\documentclass[12pt]{article}
\usepackage{amsmath,amssymb,epsfig,cite}

\normalsize

\let\oldtheequation=\theequation
\def\doteqs#1{\setcounter{equation}{0}
            \def\theequation{{#1}.\oldtheequation}}

\newcounter{sxn}
\def\sx#1{\addtocounter{sxn}{1} \bigskip\medskip \goodbreak
\noindent{\large\bf
\centerline{\thesxn.~~#1}} \nobreak \medskip}
\def\sxn#1{\sx{#1} \doteqs{\thesxn}}

\newcounter{axn}
\def\ax#1{\addtocounter{axn}{1} \bigskip\medskip \goodbreak
\noindent{\large\bf
{\Alph{axn}.~~#1}} \nobreak \medskip}
\def\axn#1{\ax{#1} \doteqs{\Alph{axn}}}

\def\br{\begin{thebibliography}{abc}}
\def\er{\end{thebibliography}}

\begin{document}
\begin{flushright}
{\sf SINP-TNP/00-05}\\
{\sf SU-4240-715}\\
\end{flushright}
\thispagestyle{empty}
\centerline{\large\bf Quantum Field Theories on Null Surfaces }
\bigskip
\begin{center}
Kumar S. Gupta$^{*}$\footnote{E-mail: gupta@tnp.saha.ernet.in},
Badis Ydri$^{+}$\footnote{E-mail: idri@suhep.phy.syr.edu}.\\
\bigskip
$^{*}${\it Saha Institute of Nuclear Physics, 1/AF Bidhannagar,\\
Calcutta - 700 064, India.}\\
\bigskip
$^{+}${\it Department of Physics , Syracuse University ,\\
Syracuse, NY 13244-1130, U.S.A.}\\ 
\end{center}

\vskip.5cm
\begin{abstract}
We study the behaviour of quantum field theories defined on a surface $S$
as it tends to a null surface $S_n$. In the case of a real, free scalar
field theory the above limiting procedure reduces the system to one with a
finite number of degrees of freedom. This system is shown to admit a one 
parameter family of inequivalent quantizations. A duality symmetry present 
in the model can be used to remove the quantum ambiguity at the self-dual 
point . In the case of the non-linear $\sigma$-model with the 
Wess-Zumino-Witten term a similar limiting behaviour is obtained. The 
quantization ambiguity in this case however cannot be removed by any means.

\end{abstract}
\newpage
\setcounter{page}{1}
\sxn{Introduction}

Quantum Field Theory (QFT) on null surfaces have been studied in different
contexts for a long time.
One example of such theories consists of
QFT on the light cone \cite{dirac}. Analysis of these
systems has led to the discovery of a rich underlying structure of field
theories and gauge theories on null surfaces \cite{hari}.
Another system of interest in this context involves the dynamics of the
degrees of freedom on the horizon of a black hole, 
the  horizon being a null surface.
Analysis of the boundary field theories on black hole horizons has
recently led to a new understanding of black hole entropy and other related
issues \cite{bh}.

Suppose that a field theory is defined on a surface $S$ which is embedded in
a flat Minkowskian manifold ${\cal M}$.
Let the embedding of $S$ in ${\cal M}$ be parametrized by a quantity
$v$
whose limiting value $v_{n}$ corresponds to a null surface
 $S_n$ .  It is natural to ask the question as to how a field theory 
defined on $S$ evolves as the parameter $v$ varies. In particular,
a field theory defined  on a null surface $S_n$ can be 
thought of as a limit of the theory defined on $S$ as 
$v  \rightarrow v_n$. It is this limiting case that we propose
to investigate in this paper.

The metric $h$ induced on $S$ from ${\cal M}$ is a function of 
the parameter $v$. When $v$ is
away from its limiting value $v_n$, 
the induced metric on $S$ is well defined.
However, as $v  \rightarrow v_n$
the induced metric $h$ tends to become degenerate. 
Correspondingly, any regular metric based action
defined on a surface $S$ fails to have a well defined limit
as $S$ tends to $S_n$ . 

A physical example of where this scenario may occur 
can be described as follows.
Considering a gravitationally collapsing sperical shell $S$ 
of dust on which some field theory
is defined. The collapsing surface $S$ at any stage can be parametrised by 
a quantity $v$. If the situation is such that the system eventually
tends to a black hole, the surface $S$ finally would tend to a null surface
$S_n$. Field theories on the surface of such a black hole could be studied
using the above mentioned limiting procedure. 

In this paper we analyze the behaviour of field theories on a 
surface $S$ as $S \rightarrow S_n$.
The analysis  is based on specific examples where 
all the issues involved can be seen in an explicit
fashion. In Section 2 we study the case of an 
abelian, free scalar field theory.
In the limiting case this model reduces to a system with finite number of
degrees of freedom that admits a one parameter family of inequivalent
quantizations. This model also exhibits a type of duality symmetry. 
The quantization ambiguity can be removed if the system is at the self-dual
point.
Section 3 describes the analysis as applied to the  $SU(l)$  
Wess-Zumino-Witten (WZW) model. The parameter in front of the action in this
case is constrained from topological considerations. 
In the limit of $S \rightarrow S_n$, 
the  quantum theory in this case is  described by a finite
 degrees of freedom and is characterized by an arbitrary
 parameter just as in the scalar field theory . We
 conclude the paper in Section 4 with a summary and outlook.

\sxn{Scalar Field}

Let ${\cal M}$ be a flat Minkowskian manifold in 2+1 dimensions 
whose spatial slice has the topology of a cylinder $ S^1 {\times} R $ . 
Let $r$ be the radius of $S^1$ and let $\theta$ be the angle spanning it.
Consider the following flat metric in ${\cal M}$ given by
\begin{equation}
ds^{2}=dt^{2}-dz^{2}-r^2d{\theta}^{2},
\end{equation}
Let $S$ given by $z=vt$ be a surface embedded in ${\cal M}$ where $v$ is the
parameter defining the embedding. The limiting value of this parameter is
given by $v = 1$.
The pull-back of the 
 above metric in ${\cal M}$ 
to the time-like surface $S$ can be written as 
\begin{equation}
ds^2|_{z=vt}=(1-v^2)dt^2-r^2d{\theta}^2 = h_{ab}dy^{a}dy^{b}.
\end{equation}
As $v \rightarrow  1$, the surface $S$ tends to 
 the null surface $S_n$ given by $z = t$ .
From Eqn. (2.2) it is easilly seen that
 the  metric $h_{ab}$ inudced on $S$ is degenerate as 
$S \rightarrow S_n$.

Consider a single real scalar field $\phi$ which is defined on the surface 
$S$. We will assume that the scalar field is valued  in a circle.
As we shall see later, the degeneracy of the metric in the limit of 
$v \rightarrow  1$ leads to a Hamiltonian that is ill-defined. In order to
address this problem we consider a renormalized field 
$ f \phi $ where $f$ is the renormalization parameter.
The action for such a real scalar field can be written as 
\begin{equation}
{\cal S}={\int}_S \sqrt{h}d^2y~ {\cal L}=\frac{f^2}{8{\pi}}
{\int}_S \sqrt{h}d^2y h^{ab} {\partial}_a{\phi} {\partial}_b{\phi}.
\end{equation}
As is evident from Eqn. (2.3), $f$ can also be interpreted as the coupling
constant of this model. 

The field $\phi$ obeys the equation of motion  
\begin{equation}
h^{ab}{\partial}_{a}{\partial}_{b}{\phi}=0.
\end{equation}
In terms of a variable $x = r{\gamma}{\theta}$ 
where ${\gamma}=\frac{1}{\sqrt{1 - v^2}}$ , 
the above action has the form
\begin{eqnarray}
{\cal S}&=&\int dt \int_{0}^{x_0} {\cal L}\nonumber\\
&=&\frac{f^2}{8{\pi}}\int dt \int_{0}^{x_0} dx 
[ ({\partial}_{t}{\phi})^2 -({\partial}_{x}{\phi})^2 ].\nonumber\\ 
\end{eqnarray}
where $x_0=2{\pi}r{\gamma}$
is the period of the  variable $x$. 
In terms of the variables $x$ and $t$ the action ${\cal S}$ 
is that of a free scalar field in 1+1 dimensions with a
diagonal metric of signature (1,-1) . 
The equation of motion following from the action ${\cal S}$ is 
\begin{equation}
[({\partial}_{t})^2-({\partial}_{x})^2]{\phi}=0.
\end{equation}

\vskip 5mm
\noindent
{\bf 2.1 Canonical Quantization}
\vskip 5mm
\noindent
The mode expansion for the real 
field $\phi$ defined on a circle has the form
\begin{equation}
{\phi}(t,x)=\phi_0 +  {\phi}_{\rm {osc}}(t,x)
\end{equation}
where
\begin{equation}
\phi_0 =  Q +\frac{N}{r{\gamma}} x + {\frac{2P}{f^2}}t,
\end{equation}
and
\begin{eqnarray}
{\phi}_{\rm {osc}}(t,x)&=&\frac{1}{f} \sum_{k>0}
[ \frac{A_{k}}{\sqrt{k}}e^{-ikx_{+}} +\frac{{A_{k}}^{+}}
{\sqrt{k}}e^{ikx_{+}}\nonumber\\
&&+\frac{B_{k}}{\sqrt{k}}e^{-ikx_{-}} + 
\frac{{B_{k}}^{+}}{\sqrt{k}}e^{ikx_{-}}].\nonumber\\
\end{eqnarray}
In Eqn. (2.9) $x_{\pm}=t{\pm}x $ and $ B_k=A_{-k} $ . 

As mentioned before, the field $\phi$ is assuemd to be valued in a circle
for all time $t$.
It therefore satisfies the consistency condition
\begin{equation}
{\phi}_0(t,x+ 2 \pi r \gamma)={\phi}_0(t,x) + 2{\pi}m,
\end{equation}  
where m is an integer. 
From the above mode expansion of the field $\phi$ and Eqn. (2.10)
it follows that
\begin{equation}
 k r \gamma = n~~~~~~\rm{(n~ is~ an~ integer),}
\end{equation}
and
\begin{equation}
N = m.
\end{equation}

The canonical momentum conjugate to the field $\phi$ is defined by 
\begin{equation}
{\pi}(t,x)=\frac{f^2}{4{\pi}}{\partial}_{t}{\phi}.
\end{equation}
Using Eqns. (7), (8) and (13)  we get that
\begin{equation}
\pi(t,x) = \frac{f^2}{4{\pi}}{\partial}_{t}{\phi}_{\rm {osc}}
 + \frac{P}{2{\pi}}.
\end{equation}

In the quantum theory, 
the wave-functional $\psi$ is a function of the field $\phi$.
Since $\phi(x)$ is identified with $\phi(x) + 2 \pi$, the $\phi_0$
dependency of the wave-functional $\psi$ satisfies the condition
\begin{equation}
\psi (\phi_0 + 2 \pi) = {\rm e}^{i 2 \pi \alpha} \psi (\phi_0),
\end{equation}
where $\alpha$ is a real number between 0 and 1. Since we are dealing with
bosonic variables alone, it is natural to choose $\alpha = 0$ corresponding
to periodic boundary condition. It therefore follows from Eqn. (2.15) that
\begin{equation}
\psi_m (\phi_0) = {\rm e}^{i p \phi_0},~~~ {\rm (~p~ is~ an~ integer~)}
\end{equation}
are the eigenfunctions of the operator 
\begin{equation}
P = \int dx \pi (x) = -i \frac{\partial}{\partial \phi_0}
\end{equation}
and the corresponding eigenvalues are $p$.
The spectrum of the operator $P$ therefore consists of integers $p$.

The canonical commutaion relations of the basic field variables are given by
\begin{equation}
[{\phi}(t,x),{\pi}(t,y)]=i{\delta}(x-y)
\end{equation}
and
\begin{equation}
[{\phi}(t,x),{\phi}(t,y)]=[{\pi}(t,x),{\pi}(t,y)]=0.
\end{equation}
It follows that
\begin{equation}
[Q,P]=i,
\end{equation}
\begin{equation}
[A_k,{A_k^{'}}^{+}]={\delta}_{k k^{'}}
\end{equation}
and all other commutation relations are zero .

\vskip 5mm
\noindent
{\bf 2.2 Ground State Energy}
\vskip 5mm
\noindent
The Hamiltonian of the system is given by
\begin{eqnarray}
H&=&\int_{0}^{x_0} dx [ {\pi}(t,x){\partial}_{t}{\phi}(t,x) -
{\cal L} ]\nonumber\\
&=&\frac{f^2}{8{\pi}}\int_{0}^{x_0} dx 
[ ({\partial}_{t}{\phi})^2 +({\partial}_{x}{\phi})^2]\nonumber\\
&=&H_0 + H_{\rm osc} ,\nonumber\\
\end{eqnarray}
where $H_0$ and $H_{\rm osc}$ are the 
Hamiltonians for the zero (or winding) and oscillating
modes respectively. They are given by
\begin{equation}    
H_0= \frac{r{\gamma}}{f^2} P^2 + \frac{f^2}{4r{\gamma}} m^2
\end{equation} 
and
\begin{equation}
H_{\rm osc}=r{\gamma} \sum_{k{\neq}0} |k| [ A_k^{+}A_{k} 
+\frac{1}{2}]
\end{equation}


A given zero mode sector is characterized by the integers
$p$ and $m$. Let the ground state (or the vacuum) in this sector 
be denoted by $|0>_{pm}$. The vacuum satisfies the
condition
\begin{equation}
H_{\rm osc}|0{\rangle}_{pm} = 0
\end{equation}
where in the above $H_{\rm osc}$ is assumed to have been normal ordered and
the zero-point energy has been subtracted. 
The ground state energy in a given zero mode sector 
characterised by the integers $p$ and $m$ satisfies the equation
\begin{equation}
H|0 \rangle_{pm}= (H_0 + H_{\rm osc})|0 {\rangle}_{pm} 
= E_G|0 {\rangle}_{pm}
\end{equation}
where
\begin{equation}
E_G=\frac{r{\gamma}}{f^2} p^2 + \frac{f^2}{4r{\gamma}}m^2
\end{equation}
and  p is the eigenvalue of the operator P in the ground state under
consideration.

Let us now define the quantity ${\tilde r}$ by
\begin{equation}
{\tilde r}=\frac{2{\gamma}r}{f^2}
\end{equation}
which is an effective radius for the system.
Then the ground state energy can be written as
\begin{equation}
E_G=\frac{1}{2}[{\tilde r}p^2 + \frac{1}{{\tilde r}}m^2].
\end{equation}

In  the limit when $v \rightarrow 1$, the induced metric $h_{ab}$ in
Eqn. (2.2)
tends to blow up. In this limit ${\tilde r}$ and the ground state
energy $E_G$ also become undefined. It may thus seem that there is now
smooth way of taking the aforementioned limit. 

We can however use the following ``renormalization group inspired"
prescription to make this limit well defined. Let us first note that
 the quantity $v$ used to define the embedding of $S$ in ${\cal M}$ 
 could be thought as a regulator.
We are really interested in the situation $S \rightarrow S_n$ or
$v \rightarrow 1$, which can be thought of as the regulator being removed.   
In order for the ground state energy to have a
smooth behaviour in this limit we postulate that               
 the coupling constant $f$ is  a function of the regulator $v$.
The functional dependence of $f$ on $v$ is to be determined from the
physical condition that as $v \rightarrow 1$,
the ground state energy should
be independent of $v$. In other words, as 
$v \rightarrow 1$
\begin{equation}
\frac{dE_G}{dv}=0.
\end{equation}
Using the equation $(2.27)$ we find that

\begin{equation}
\frac{dE_G}{dv}=\frac{1}{2{\tilde r}}\frac{d {\tilde r}}{dv}({\tilde r}p^2 
- \frac{1}{{\tilde r}}m^2)
\end{equation}
The second term cannot be zero for all values of
$v$ because $p$ and $m$ are fixed numbers . We are hence left with
the condition
\begin{equation}
\frac{1}{2{\tilde r}}\frac{d {\tilde r}}{dv}=0
\end{equation}
This is possible only if in the limit $v \rightarrow 1$, 
\begin{equation}
f(v)=F\sqrt{\gamma}
\end{equation}
where $F$ is a constant parameter and can be thought of as the
``renormalized" coupling constant.

Using Eqns. (28), (29) and (33), the ground state
 energy has the form
\begin{equation}
E_G=\frac{r}{F^2}p^2 + \frac{F^2}{4r}m^2.
\end{equation}
In any given zero-mode sector,
different choices of the parameter $F$ 
would lead to different values of the
ground state energy and hence to inequivalent quantum field theories. 
The parameter $F$ is not determined by the above analysis
 and can presumably
be obtained from empirical considerations.

It should be noted that only the zero mode sector has information about the
coupling constant $f$ and consequently of the parameter $F$. 
The expression for 
$H_{\rm osc}$ (cf. Eqn. (2.24)) is independent of $f$.
As  $v \rightarrow 1$ , we see from Eqn. (2.11) that for any given $n$,
$k \rightarrow 0$. However, from Eqn. (2.24), only nonzero $k$ contributes
to $H_{\rm{osc}}$. Hence the contribution to the Hamiltonian coming from the 
the oscillatory modes
become energetically unfavourable as $S \rightarrow S_n$.
We therefore arrive at the conclusion that as $S \rightarrow S_n$, the
model under consideration reduces to a  quantum mechanical system with a
finite number of degrees of freedom given by the zero modes of the
original problem. The
Hamiltonian of this reduced system is still given
 by Eqn. (2.23) where $f$ is given by Eqn. (2.33) and the corresponding
eigenavlues are given by Eqn. (2.34).

\vskip 5mm
\noindent
{\bf 2.3 Duality}
\vskip 5 mm
\noindent
A real scalar field system valued in a circle has a well known duality
symmetry \cite{dual}.
A remnant of that can be seen in  the reduced system obtained above.
For a fixed $r$, our system is characterized by the 
numbers $p$, $m$ and $F$.  Under the transformations
\begin{eqnarray}
p&{\longrightarrow}&m\nonumber\\
m&{\longrightarrow}&p\nonumber\\
\frac{r}{F^2}&{\longrightarrow}&\frac{F^2}{4 r}
\end{eqnarray}
$E_G$ belonging to two different configurations get interchanged. This is
analogous to a T-duality.

The duality symmetry by itself imposes no restriction on $F$.
 There is however a special configuration,
namely the ``self-dual" point where the duality symmetry can be used to
fix the arbitrariness in $F$. The ``self-dual" point is given by
\begin{equation}
\frac{r}{F^2} = \frac{F^2}{4 r}.
\end{equation}
At this special point $F$ therefore satisfies the condition
\begin{equation}
F^2 = 2 r.
\end{equation}
Using Eqns. (34) and (37), the ground state energy can them be expressed as 
\begin{equation}
E_G=\frac{1}{2}[p^2 + m^2]
\end{equation}
which is independent of $F$.
The limiting procedure described above thus leads to a unique 
quantization only at the self-dual point.

\sxn{Non Linear ${\sigma}$ - Model}

Let ${\cal C} $ be a three dimensional manifold whose boundary 
$ {{\partial}{\cal C}} $ is a two dimensional Minkowskian manifold with
a topology of $S^1 {\rm x} R$. We identify $ {{\partial}{\cal C}} $ 
with a surface $S$ on which the induced metric $h^{ab}$ is given by Eqn(2.2) . 
The action for the non linear $\sigma$-model with a
Wess-Zumino-Witten term (WZW) \cite{witten}  for a group G is given by
\begin{eqnarray}
S&=&S_0+S_{\rm wzw},\nonumber\\
S_0&=&A\int_{{\partial}{\cal C}} d^2y 
{\sqrt{h}}h^{{\mu}{\nu}}Tr{\partial}_{\mu} g {\partial}_{\nu}g^{-1},\nonumber\\
S_{\rm wzw}&=&\int_{\cal C}{\Omega}=B \int_{\cal C} Tr(g^{-1}dg)^3,
\end{eqnarray} 
where $A$ and $B$ are constants and $g$ takes values in the group $G$.
For simplicity we will assume in this section that $G=SU(l)$ with $l>1$ . 
The coefficient $B$ of the second term in the action is not arbitrary. 
From topological considerations $B = \frac{n}{24 \pi}$ ,
where $n$ is an integer \cite{witten}.

Let $ T_{a},a=1,..r $ be the generators of $G$
satisfying the comutation relations
\begin{equation}
[T_a,T_b]=if_{abc}T_c .
\end{equation}
These generators are normalised in such a way that 
\begin{equation}
TrT_aT_b=2{\delta}_{ab}.
\end{equation}
The group $G$ can act on $g$ either from the 
left or from the right . The left action is given by 
\begin{equation}
g{\longrightarrow}g^{'}=g + i g x_a T_a .
\end{equation}
where $ x_a $ are small parameters . 
The right action of the group is similarly given by 
\begin{equation}
g{\longrightarrow}g^{'}=g + i x_a T_a g .
\end{equation} 
For both cases, the variation of $ S $  is  
\begin{eqnarray}
{\delta}S&=&{\delta}S_0 + {\delta}S_{wzw},\nonumber\\
{\rm where}~~
{\delta}S_0&=&2A\int d^2x {\sqrt{h}}h^{{\mu}{\nu}} 
Trg^{-1}{\delta}g {\partial}_{\mu}(g^{-1}{\partial}_{\nu}g) 
+ \rm{~Total ~derivatives }\nonumber\\ 
{\rm and}~~
{\delta}S_{wzw}&=&-3B\int d^2x {\epsilon}^{{\mu}{\nu}} 
Trg^{-1}{\delta}g{\partial}_{\mu}(g^{-1}{\partial}_{\nu} g ).\nonumber\\
\end{eqnarray}
By setting $ {\delta}S=0 $ we get as equations of motion 
\begin{equation}
2A{\sqrt{h}}h^{{\mu}{\nu}}{\partial}_{\mu}(g^{-1}{\partial}_{\nu} g)-3B
{\epsilon}^{{\mu}{\nu}}{\partial}_{\mu}(g^{-1}{\partial}_{\nu} g)=0.
\end{equation}
In terms of the light cone coordinates
\begin{equation}
x_{\pm}=t{\pm}x,
\end{equation}
Eqn. (3.7) can be expressed as  
\begin{equation}
(2A -3B){\partial}_{+}(g^{-1}{\partial}_{-}g)+
(2A +3B){\partial}_{-}(g^{-1}{\partial}_{+}g)=0.
\end{equation}
For the choice of 
$A={\pm}{\frac{3B}{2}}$ the left and right movers decouple from each other
and the equations of motion reduces to
\begin{equation}
{\partial}_{\mp}(g^{-1}{\partial}_{\pm}g)=0.
\end{equation} 
The  currents arising  from the left action of the
group are given by 
\begin{eqnarray}
J^{\mu}_a&=&h^{{\mu}{\nu}} Tr T_ag^{-1}{\partial}_{\nu}g,\nonumber\\
{\rm with}~~
J^{0}_a&=&{\gamma}^2TrT_a g^{-1}\dot{g}\nonumber\\
{\rm and}~~
J^{1}_a&=&-\frac{1}{r^2}TrT_a g^{-1} g^{'},\nonumber\\
\end{eqnarray}
where $ \dot{g}={\partial}_{t}g $ and $ g^{'}={\partial}_{\theta}g $.
Using Eqn. (3.11) the light-cone components of the currents can be written
as 
\begin{eqnarray}
J^{\pm}_{a}&=&Tr T_a g^{-1}{\partial}_{\pm}g\nonumber\\
&=&\frac{1}{2{\gamma}^2}J^{0}_a {\mp}\frac{r}{2{\gamma}}J^{1}_a. 
\end{eqnarray}

The currents resulting from the right action of the group can also be
 found in a similar fashion. They are 
\begin{eqnarray}
\bar{J}^{\pm}_{a}&=&Tr T_a ({\partial}_{\pm}g)g^{-1}\nonumber\\
&=&\frac{1}{2{\gamma}^2}\bar{J}^{0}_a {\mp}\frac{r}{2{\gamma}}\bar{J}^{1}_a,
\end{eqnarray}
where
\begin{eqnarray}
\bar{J}^{0}_a&=&{\gamma}^2TrT_a\dot{g}g^{-1}\nonumber\\
\bar{J}^{1}_a&=&-\frac{1}{r^2}TrT_a g^{'}g^{-1}.
\end{eqnarray}
\vskip 5mm
\noindent
{\bf 3.1 The Canonical Formalism}
\vskip 5mm
\noindent
Let $ {\xi}_i, ~i=1,...,{\rm dim}~G$ 
be a set of local coordinates parametrizing the elements $g
\in G$\cite{bal}.In terms of these local coordinates $S_0$ can be 
 expressed as 
\begin{eqnarray}    
S_0&=&\int d^2x {\cal L}_0,\nonumber\\
{\cal L}_0&=&Ar{\gamma}Tr\frac{{\partial}g}{{\partial}{\xi}_i}
\frac{{\partial}g^{-1}}{{\partial}{\xi}_j}[ \dot{{\xi}_i}\dot{{\xi}_j} 
-\frac{1}{r^2{\gamma}^2}
{\xi}^{'}_i {\xi}^{'}_j ].
\end{eqnarray}
We would also like to express the WZW part of the action in terms of the
coordinates $\xi$. The WZW terms cannot be written globally in terms of a
single set of local coordinates. To proceed we 
assume that the group manifold
consists of a number of patches labelled by a parameter $u$. 
The restriction of 
${\Omega}=BTr(g^{-1}dg)^3$ to any of these patches can be written as 
\begin{equation}
{\Omega}^{u}=d{\omega}^{u},
\end{equation}
where ${\omega}^{u}$ is  a two form defined by 
\begin{equation}
{\omega}^{u}=\frac{1}{2}{\omega}^{u}_{ij} d{\xi}^{i}{\wedge}d{\xi}^{j}.
\end{equation}
From Eqns. (3.16) and (3.17),  $\Omega^u$ has the form
\begin{equation}
{\Omega}^{u}=\frac{1}{6}{\Omega}^{u}_{ijk}d{\xi}^{i}{\wedge}d{\xi}^{j}
{\wedge}d{\xi}^{k},
\end{equation}
where ${\Omega}^{u}_{ijk}$ is given by
\begin{equation}
{\Omega}^{u}_{ijk}=\frac{{\partial}{\omega}^{u}_{jk}}{{\partial}{\xi}^{i}} 
+ \frac{{\partial}{\omega}^{u}_{ki}}{{\partial}{\xi}^{j}} 
+ \frac{{\partial}{\omega}^{u}_{ij}}{{\partial}{\xi}^{k}}.
\end{equation}
${\Omega}^{u}_{ijk}$ can also be expressed however
 in terms of the group element $g$
as 
\begin{equation}
{\Omega}^{u}_{ijk}=3BTr[g^{-1}{\frac{{\partial}g}
{{\partial}{\xi}^{i}}},g^{-1}{\frac{{\partial}g}
{{\partial}{\xi}^{j}}}]g^{-1}{\frac{{\partial}g}{{\partial}{\xi}^{k}}}.
\end{equation}
In terms of the local coordinates the WZW action then takes the form
\begin{eqnarray}
S_{wzw}&=&\sum_{u}\int_{{\partial}{\cal C}}{\cal L}^{u}_{wzw}d^2x,\nonumber\\
{\rm where}~~
{\cal L}^{u}_{wzw}&=&\frac{1}{2}{\omega}^{u}_{ij}
{\partial}_{a}{\xi}^{i}{\partial}_{b}{\xi}^{j}{\epsilon}^{ab}.\nonumber\\
\end{eqnarray}

We would next like to compute the canonical momentum $P_i$ conjugate to the
coordinate ${\xi}_i$.
Using Eqns. (3.15) and (3.21) $P_i$ can be written as 
\begin{eqnarray}
P_i&=&\frac{{\partial}{\cal L}_0}{{\partial}{{\dot{\xi}}_i}}+
\frac{{\partial}{\cal L}_{wzw}}{{\partial}{{\dot{\xi}}_i}}\nonumber\\
&=&Ar{\gamma}Tr[\frac{{\partial}g}{{\partial}{\xi}_i}
\frac{{\partial}g^{-1}}{{\partial}{\xi}_j}\dot{{\xi}_j} 
+\frac{{\partial}g}{{\partial}{\xi}_j}
\frac{{\partial}g^{-1}}{{\partial}{\xi}_i}\dot{{\xi}_j}] -
{\omega}^{u}_{ij}{\partial}_{\theta}{\xi}^{j}\nonumber\\
&=&-2Ar{\gamma}Trg^{-1}\dot{g}g^{-1}
\frac{{\partial}g}{{\partial}{x^{i}}}-{\omega}^{u}_{ij}
{\partial}_{\theta}{\xi}^{j}.\nonumber\\
\end{eqnarray}

The Hamiltonian density therefore has the form
\begin{eqnarray}
{\cal H}&=&P_i{\xi}_i -{\cal L}_0 -{\cal L}_{wzw}\nonumber\\
&=&-Ar{\gamma}Tr [ (\dot{g}g^{-1})^2 
+\frac{1}{r^2{\gamma}^2}({g^{'}}g^{-1})^{2}].\nonumber\\
\end{eqnarray}
As expected, there is no contribution from the WZW term to the Hamiltonian. 
In terms of the light cone currents defined in Eqn. (3.12) ,
the Hamiltonian density can be written as
\begin{equation}
{\cal H}=-Ar{\gamma}[{J_a^{+}}^2 +{J_a^{-}}^2].
\end{equation}

Next we turn to the commutation relations for this system.
The basic commutation relations  are given by
\begin{equation}
[{\xi}_a(\theta),{\xi}_b({\theta}^{'})]=[P_a({\theta}),P_b({\theta}^{'})]=0 
\end{equation}
and
\begin{equation}
[{\xi}_a(\theta),P_b({\theta}^{'})]=
i{\delta}_{ab}{\delta}(\theta - {\theta}^{'})
\end{equation}
From these commutation relations it follows that
\begin{equation}
[g(\theta),P_b({\theta}^{'})]=i\frac{{\partial}{g(\theta)}}{{\partial} 
{\xi}_b}{\delta}({\theta} - {\theta}^{'}).
\end{equation}
Similarly we have
\begin{equation}
[g(\theta)^{-1},P_b({\theta}^{'})]=i\frac{{\partial}{g(\theta)^{-1}}}
{{\partial}{\xi}_b}{\delta}({\theta} - {\theta}^{'}).
\end{equation} 
It is however useful to rewrite 
these commutation relations in a  form that is independent of the local
coordinates $\xi$ \cite{bal}.
 To this end,
we  introduce a new set of functions 
${\xi}(x)$ with the condition that
$ {\xi}(0)={\xi} $ . The field $ g({\xi})$ can
 be understood as the value of a field $ g({\xi}(x)) $ at $ {\xi}(0)$ . The
 field $ g({\xi}(x)) $ is defined by
\begin{equation}
g({\xi}(x))=g({\xi}(0)) {\rm exp} (ix_a T_a)
\end{equation}
Differentiating with respect to $ x_a $ and then setting $ x=0 $
 we get the  identity
\begin{equation}
N^{a}_{b} \frac{{\partial}g}{{\partial}{\xi}^a} = i g({\xi}) T_b
\end{equation}
where $ N^{a}_{b} =\frac{{\partial}{\xi}^a}{{\partial}x_b}{|}_{x=0} $
can be proven to be nondegenerate \cite{bal}. Using Eqn. (3.30),
we can replace the phase space 
variables $ P_a $ with new variables
$ {\Pi}_a $ defined as    
\begin{eqnarray}
{\Pi}_a&=&N^{b}_{a}P_{b}\nonumber\\
&=&-2i\frac{Ar}{\gamma} J^{0}_{a} -N_{a}^{b}
{\omega}^{u}_{bc}{\partial}_{\theta}{\xi}_c.\nonumber\\
\end{eqnarray}
Using Eqns. (3.27), (3.28) ,(3.30) and $(3.31)$ it  follows that
\begin{equation}
[g(\theta),{\Pi}_b({\theta}^{'})]=-{\delta}({\theta} - {\theta}^{'})g(\theta)
T_b
\end{equation}
and
\begin{equation}
[g^{-1}(\theta),{\Pi}_b({\theta}^{'})]=
{\delta}({\theta} - {\theta}^{'})T_b g^{-1}(\theta).
\end{equation}
The above commutators between $g$, $g^{-1}$ and $\Pi$ carry no explicit
dependence on the local coordinates $\xi$.
Finally, the commutator of the $\Pi$'s is 
given by (see the appendix for the proof)
\begin{equation}
[{\Pi}_a({\theta}),{\Pi}_b({\theta}^{'})]
=i f_{abc}{\Pi}_c({\theta}) {\delta}(\theta -{\theta}^{'})
\end{equation}
and all other commutation relations are trivial.
Eqns (3.32), (3.33) and (3.34) 
embody the fundamental commutators for  this system.

\newpage
\noindent
{\bf 3.2 Current Algebra}
\vskip 5mm
\noindent
We are now ready to calculate the current commutators. First we note that
 the expression for $J^{1}_a(\theta)$ contains no time derivative. It
therefore follows that
\begin{equation}
[J^{1}_a(\theta),J^{1}_b({\theta}^{'}]=0.
\end{equation}
Next, by differentiating Eqn. (3.32) with respect to $\theta$ we get
\begin{equation}
[{\partial}_{\theta}g(\theta),{\Pi}_b({\theta}^{'})]=
-{\delta}({\theta} - {\theta}^{'}){\partial}_{\theta}g({\theta})T_b - 
{\partial}_{\theta}{\delta}({\theta} - {\theta}^{'})g(\theta)T_b.
\end{equation}
Using the above relation and the definition of the
 current $J_a^{1}$ (cf. Eqn. (3.11)), it can be shown that 
\begin{equation}
[{\Pi}_a(\theta),J^{1}_{b}({\theta}^{'})]=
if_{abc} J^{1}_{c}(\theta) + 
\frac{TrT_a T_b}{r^2} {\partial}_{\theta}{\delta}(\theta -{\theta}^{'}).
\end{equation}
From Eqn. (3.31) we have 
$ J_a^0=i{\gamma}/2Ar[{\Pi}_a + 
N_a^{b}{\omega}_{bc}^{u}{\partial}_{\theta}{\xi}_c] $.
Using this expression for $ J_a^0$ and Eqn. (3.37) we get the second current
commutator as 
\begin{equation}
[J^{0}_a({\theta}),J^{1}_b({\theta}^{'})]=-\frac{\gamma}{2Ar} f_{abc} 
J^{1}_{c}(\theta) {\delta}({\theta} - {\theta}^{'}) +
 \frac{i{\gamma}}{2Ar^3} TrT_a T_b {\partial}_{\theta}{\delta}({\theta} -
{\theta}^{'}).
\end{equation}
Finally, a  tedious calculation (the details are shown in the appendix) 
 gives  the last  current commutator as
\begin{equation}
[J^{0}_a({\theta}),J^{0}_b({\theta}^{'})]=-\frac{\gamma}{2Ar} f_{abc} 
J^{0}_{c}(\theta) {\delta}(\theta - {\theta}^{'}) 
- \frac{3B{\gamma}^2}{4A^2} {\delta}({\theta} 
-{\theta}^{'}) f_{abc}J^{1}_{c}(\theta)
\end{equation}

We would next like to obtain the  commutators for the light-cone
components of the currents . Using Eqns. 
(3.35), (3.38) and (3.39) we get the following currents algebra
\begin{eqnarray}
[ J^{+}_a(\theta) , J^{+}_b({\theta}^{'}) ]&=&-\frac{1}{8Ar{\gamma}^3} f_{abc} 
{\delta}({\theta} -{\theta}^{'}) J^{0}_c(\theta)\nonumber\\
&+& \frac{1}{4A{\gamma}^2} (1 -
\frac{3B}{4A}) f_{abc}J_{c}^{1}(\theta)
{\delta}(\theta-{\theta}^{'})\nonumber\\
&+&\frac{1}{4iAr^2{\gamma}^2}TrT_aT_b 
{\partial}_{\theta}{\delta}({\theta} -{\theta}^{'}).\nonumber\\
\end{eqnarray}
Let us now proceed by  considering the two cases 
$A={\frac{3B}{2}}$ and $A=-{\frac{3B}{2}}$ separately .
For the first case where $ \frac{3B}{2A}=1, $ Eqn. (3.40)
reduces to the Kac-Moody algebra 
\begin{equation}
[ J^{+}_a(\theta) , J^{+}_b({\theta}^{'}) ]=-\frac{1}{4Ar{\gamma}} f_{abc} 
{\delta}({\theta} -{\theta}^{'}) J^{+}_{c} 
- \frac{i}{2Ar^2{\gamma}^2}{\delta}_{ab}{\partial}_{\theta}{\delta}({\theta} 
-{\theta}^{'}) .
\end{equation}

The currents ${\bar{J}}^{-}_{a}=TrT_a{\partial}_{-}g g^{-1} $ 
coming from the right action of the group would similarly generate another
Kac-Moody algebra. To get the
 algebra generated by  ${\bar{J}}^{-}_{a}$ first note 
that the action is invariant under
 the transformations  
${\theta}{\longrightarrow}-{\theta}$ and 
$ g{\longrightarrow}g^{-1} $ . 
Under these transformations $J^{+}_a{\longrightarrow}-\bar{J}^{-}_a$ 
and therefore the current commutator in   Eqn. (3.41) becomes
\begin{equation}
[ {\bar{J}}^{-}_a(\theta) , {\bar{J}}^{-}_b({\theta}^{'}) ]=
\frac{1}{4Ar{\gamma}} f_{abc} {\delta}({\theta} 
-{\theta}^{'}) {\bar{J}}^{-}_{c} +\frac{i}{2Ar^2{\gamma}^2}
{\delta}_{ab}{\partial}_{\theta}{\delta}({\theta} -{\theta}^{'}) .
\end{equation} 
We also have 
\begin{equation}
[J^{+}_{a}(\theta),{\bar{J}}^{-}_{b}({\theta}^{'})]=0
\end{equation}
as the two currents come from the two commuting 
actions of the group on itself .

For the second point $\frac{3B}{2A}=-1$ 
exactly the same arguments will 
lead to the two other commuting 
Kac-Moody algebras given by the currents
 $J_{a}^{-}$ and $\bar{J}_{a}^{+}$ . The first
 currents algebra generated by $J_a^{-}$ has the form $(3.41)$ with
 the substitution $J_{a}^{+}{\longrightarrow}-J_{a}^{-}$ and 
$A{\longrightarrow}-A$ . The currents algebra corresponding
 to $\bar{J}_{a}^{+}$ is aobtained from $(3.42)$ by
 a similar substitution $\bar{J}_{a}^{-}{\longrightarrow}-\bar{J}_{a}^{+}$ 
and $A{\longrightarrow}-A$ .

\vskip 5mm
\noindent
{\bf 3.3 Mode Expansion}
\vskip 5mm
\noindent

Let us  consider  the case when $\frac{3B}{2A}=1$ 
(the treatement of the case $\frac{3B}{2A}=-1$ is exactly similar). 
We first express the the Hamiltonian density ${\cal H}$ in
terms of the two commuting
set of currents $J^{+}_{a}$ and ${\bar{J}}^{-}_{a}$ which are relevant to
the case under consideration. To this end we note that a given element $L$
in the Lie algebra of $G$ can be written as
$ L=\frac{T_a}{2}Tr(T_a L) $ 
Using this we  can then check that $ {J^{-}_{a}}^2=({\bar{J}}^{-}_{a})^2$.
The Hamiltonian density in Eqn. (3.24) can then be expressed as  
\begin{eqnarray}
{\cal H}&=&-Ar{\gamma}[(J^{+}_{a})^2 +(J^{-}_{a})^2 ]\nonumber\\
&=&-Ar{\gamma}[(J^{+}_{a})^2 +({\bar{J}}^{-}_{a})^2 ]\nonumber\\
&=&-\frac{1}{16Ar{\gamma}}[(K^{+}_{a})^2 +({\bar{K}}^{-}_{a})^2 ],
\end{eqnarray}
where $K^{+}_{a}$ and ${\bar{K}}^{-}_{a}$ are defined as 
\begin{eqnarray}
K^{+}_{a}&=&-4Ar{\gamma}J^{+}_{a},\nonumber\\
{\bar{K}}^{-}_{a}&=&4Ar{\gamma}{\bar{J}}^{-}_{a}.
\end{eqnarray}
 They  satify the commutation relations 
\begin{equation}
[ {K}^{+}_a(\theta) , {K}^{+}_b({\theta}^{'}) ]=f_{abc}{\delta}
({\theta} -{\theta}^{'}) {K}^{+}_{c}(\theta) 
-8iA{\delta}_{ab}{\partial}_{\theta}{\delta}({\theta} 
-{\theta}^{'})
\end{equation}
and
\begin{equation}
[ \bar{K}^{-}_a(\theta) , \bar{K}^{-}_b({\theta}^{'}) ]=
f_{abc}{\delta}({\theta} -{\theta}^{'}) \bar{K}^{-}_{c}(\theta) 
+8iA{\delta}_{ab}{\partial}_{\theta}{\delta}({\theta} 
-{\theta}^{'}).
\end{equation}

Next we proceed with the mode expansion of the currents. Let us first note
that in terms of $g \in G$, the current $K_{a}^{+}(x)$ has the expression 
\begin{equation}
K_{a}^{+}(x)=-2Ar{\gamma}TrT_ag^{-1}\dot{g} - 2ATrT_ag^{-1}g^{'}.
\end{equation}
A similar expression will hold for the current $\bar{K}^{-}$ . 
The mode expansion for the two terms in the rhs of Eqn (3.48) are given by
\begin{eqnarray}
TrT_ag^{-1}{\dot g}&=&\sum_{k{\neq}0}J^0_a(k)
e^{i({\omega}t - kx)} + {J^{0}_a(0)},\nonumber\\
TrT_ag^{-1}g^{'}&=&\sum_{k{\neq}0}J^{1}_{a}(k)
e^{i({\omega}t - kx)} + J^1_a(0).
\end{eqnarray}
Next we
 can check using the periodicity requirement 
\begin{equation}
J^{\mu}_{a}(t,x + x_0)=J^{\mu}_{a}(t,x) 
\end{equation}
that
\begin{equation}
kr{\gamma}=q
\end{equation}
where $q$ is an integer . Using the equations of motion (3.10), we
 see that $\omega = k$ for $J^{+}_{a}$ and $\omega = -k$ for $J^{-}_{a}$. 

Now by using Eqn. (3.49) in (3.48), we get
\begin{eqnarray}
K^{+}_{a}(x_{+})&=&\frac{1}{2i{\pi}}[\sum_{k{\neq}0}K^{+}_{a}(k)e^{-ikx_{+}}+ 
{\cal P}_a]\nonumber\\
\rm{where}~
K^{+}_{a}(k)&=&-{{4i{\pi}Ar{\gamma}}}[J^{0}_{a}(k) 
+ \frac{1}{r{\gamma}}J^{1}_{a}(k)]\nonumber\\
\rm{and}~
{\cal P}_a&=&-{{4i{\pi}Ar{\gamma}}}[J^{0}_{a}(0) +
 \frac{1}{r{\gamma}}J^{1}_{a}(0)].
\end{eqnarray}
Similarly ge wet that,
\begin{eqnarray}
\bar{K}^{-}_{a}(x_{-})&=&\frac{1}{2i{\pi}}[\sum_{k{\neq}0}
\bar{K}^{-}_{a}(k)e^{ikx_{-}} 
+ {\cal M}_a]\nonumber\\
\rm{where}~
\bar{K}^{-}_a(k)&=&{{4i{\pi}Ar{\gamma}}}[\bar{J}^{0}_{a}(k)- 
\frac{1}{r{\gamma}}\bar{J}^{1}_{a}(k)]\nonumber\\
\rm{and}~
{\cal M}_a&=&{{4i{\pi}Ar{\gamma}}}[\bar{J}^{0}_{a}(0)- 
\frac{1}{r{\gamma}}\bar{J}^{1}_{a}(0)].
\end{eqnarray}
In above,
$\bar{J}^{0}_{a}(k)$ and $\bar{J}^{1}_{a}(k)$ are 
the modes corresponding  to $TrT_a{\dot g}g^{-1}$ and 
$TrT_ag^{'}g^{-1}$ respectively .

Using the above currents and  the Kac-Moody algebra $(3.46)$ , we get that
\begin{equation}
[{\cal P}_a,{\cal P}_b]=if_{abc}{\cal P}_c
\end{equation}
and
\begin{equation}
[ K^{+}_a(p) , K^{+}_b(k) ]=if_{abc}K^{+}_{c}(p+k) + 
16{\pi}Ar{\gamma}p{\delta}_{ab}{\delta}_{p+k,0}.
\end{equation}
In the same way we get from $(3.47)$ ,
\begin{equation}
[{\cal M}_a,{\cal M}_b]=if_{abc}{\cal M}_c
\end{equation}
and
\begin{equation}
[ {\bar{K}}^{-}_a(p) , {\bar{K}}^{-}_b(k) ]=if_{abc}{\bar{K}}^{-}_{c}(p+k) - 
16{\pi}Ar{\gamma}p{\delta}_{ab}{\delta}_{p+k,0}.
\end{equation}
From $(3.54)$ and $(3.56)$ we immediately 
see that $\{{\cal P}_a\}$ and $\{{\cal M}_a\}$ 
are two representations of $SU(l)$ generators .
 
We can now compute the Hamiltonian 
\begin{equation}
H=\int_{0}^{2{\pi}} d{\theta}{\cal H}
\end{equation}
in terms of the oscillation modes $K^{+}_a(k)$ , $\bar{K}^{-}_a(k)$ and 
the zero modes ${\cal P}_a$ , ${\cal M}_a$ . The answer turns out to be 
\begin{equation}
H=H_{\rm{osc}} +H_0
\end{equation}
where
\begin{equation}
H_{\rm{osc}}=-\frac{{\pi}}{8Ar{\gamma}}\sum_{k{\neq}0}
[:K^{+}_{a}(k)K^{+}_{a}(-k): + :\bar{K}^{-}_{a}(k)\bar{K}^{-}_{a}(-k):]
\end{equation}
and
\begin{equation}
H_0=-\frac{\pi}{8Ar{\gamma}}({\cal P}^2 + {\cal M}^2).
\end{equation}
The contribution to the Hamiltonian from the oscillatory mode has been
normel ordered.
${\cal P}^2$ and ${\cal M}^2$ are simply the $SU(l)$ Casimirs
 $ {\cal P}^2=\sum_{a}{\cal P}_a^2$ and ${\cal M}^2=\sum_{a}{\cal M}_a^2$ 
respectively and are given by \cite{slanski}
\begin{eqnarray}
{\cal P}^2&=&\frac{N_{adj}}{N_{p}}p\nonumber\\ 
{\cal M}^2&=&\frac{N_{adj}}{N_m}m
\end{eqnarray}
where $N_{adj}$ is the dimension of the adjoint representation of $SU(l)$ . 
$N_p$ and $N_m$ above are the dimensions of the representaions 
$\{{\cal P}_{a}\}$ and $\{{\cal M}_{a}\}$ respectively 
and $p$ $(m)$ is  the index of the representations $\{P_a\}$ $(\{M_a\})$. 
A given zero mode will be characterized by two integers $p$ and $m$ and it 
will be denoted by $|pm \rangle $ . The state  $|pm \rangle $ will be
annihilated by $H_{\rm{osc}}$ as the latter is normal ordered.
The ground state energy of the system would therefore be given by
\begin{equation}
E_{pm}=-\frac{\pi}{8Ar{\gamma}}({\cal P}^2 + {\cal M}^2).
\end{equation}
where now ${\cal P}^2$ and ${\cal M}^2$ are being understood 
to be equal to the numbers given by the equation $(3.62)$. 
   
We want now to investigate the behaviour of the 
currents algebras and the Hamiltonian as $v \rightarrow 1$.
The currents in the equations $(3.52)$ $(3.53)$ as well as the
 Hamiltonians $(3.60)$ and $(3.61)$ are functions of the parameter
$v$ and tend to become ill defined as $v \rightarrow 1$ . As in Section 
2.2 , we can again use a
``renormalization group inspired" technique to get
 a well defined theory in this limit . The  constant $B$ in this case 
has the allowed values given by $\frac{n}{24{\pi}}$ where $n$ is an
integer. Furthermore $A$ is constrained by the condition
$A={\pm}{\frac{3B}{2}}$ . We will however 
 assume that $A$ is a function of $v$,
and its dependence on $v$ is to be determined from the condition that
in the limit of $v \rightarrow 1$, the ground state energy $E_{pm}$ becomes
independent of $v$, i.e.
\begin{equation}
\frac{dE_{pm}}{dv} = 0.
\end{equation}
However by using $(3.63)$ it immediately follows that 
\begin{equation}
\frac{{\pi}}{8r{\gamma}A}= \frac{1}{C}
\end{equation}
where $C$ is a constant. 
$A$ in the above equation is constrained as mentioned above.
It therefore cannot run continuously with $v$ and changes only in discrete
steps always satisfying the constraint. Suppose when $v = 0$, $A$ was given by
 $A_0$. The constant $C$ was then given by 
$C = \frac{8r}{\pi}A_0$. As  $v \rightarrow 1$, it follows from Eqn. (3.65)
that the limiting value of $A$ actually tends to {\it zero}. $C$ however is
finite in this limit and is 
given by the same constant value as mentioned above.

In view of the above, as $v \rightarrow 1$, the 
ground state energy of the system is given by 
\begin{equation}
E_{pm}=-\frac{1}{C}({\cal P}^2 + {\cal M}^2)
\end{equation}
 corresponding to the Hamiltonian 
\begin{equation}
H_{0}=-\frac{1}{C}({\cal P}^2 + {\cal M}^2).
\end{equation}
where now ${\cal P}_a$ and ${\cal M}_a$ are given by :
\begin{eqnarray}
{\cal P}_a&=&-\frac{i{\pi}^2 C}{2}J^{0}_a(0) \nonumber\\
{\cal M}_a&=&\frac{i{\pi}^2 C}{2}{\bar{J}}^{0}_{a}(0),
\end{eqnarray}
and they still do satisfy $(3.54)$ and $(3.56)$ respectively .
The value of the constant $C$ is related the the value of $A$ when $v = 0$.
This value is not determined by the theory and must be obtained from
empirical considerations. 
As in the scalar field case, this system also therefore admits a one
parameter family of inequivalent quantizations.

Let us now turn our attention to the oscillatory modes. 
As $v \rightarrow 1$, $A$ satisfies Eqn. (3.65) and the expressions for the
oscillatory modes are given by 
\begin{eqnarray}
K^{+}_{a}(k)&=&-\frac{i {\pi}^2 C}{2}J^{0}_{a}(k)\nonumber\\
\bar{K}^{-}_{a}(k)&=&\frac{i {\pi}^2 C}{2}\bar{J}^{0}_{a}(k).
\end{eqnarray}
The current algebra satisfied by these modes are now given by
\begin{equation}
[ K^{+}_a(p) , K^{+}_b(k) ]=if_{abc}K^{+}_{c}(p+k)
+2 {\pi}^2 C p{\delta}_{ab}{\delta}_{p+k,0}
\end{equation}
and 
\begin{equation}
[ {\bar{K}}^{-}_a(p) , {\bar{K}}^{-}_b(k) ]=if_{abc}{\bar{K}}^{-}_{c}(p+k)
-2{\pi}^2 C p{\delta}_{ab}{\delta}_{p+k,0}.
\end{equation}
However, as $v \rightarrow 1$, it is clear from Eqn. (3.51) that
$k$ must go to zero for any value of the integer $q$.
From Eqn. (3.60), we see that
the oscillatory part of the Hamiltonian has contributions only 
from those modes for which $k$ is not equal to zero. We therefore conclude
that as $v \rightarrow 1$, the oscillatory modes becomes energetically
unfavourable and do not contribute to the Hamiltonian. The entire theory in
this limit, just as in the scalar field case, 
is described by a finite number of degrees of freedom 
given only by the zero modes.

\sxn{Conclusion}

In this paper we have investigated the limitng behaviour of 
quantum field theories defined on a 
surface $S$ as the latter  tends to a null surface $S_n$.

In the case of a scalar field theory the above limiting procedure reveals
several interesting features. First, 
as $S \rightarrow S_n$, the excitation of the 
oscillatory degrees of freedom of the system becomes energetically
unfavourable. 
In this situation, the model reduces to a
quantum mechanical system with the winding modes as the only degrees of
freedom. Second, in the limit when $S \rightarrow S_n$, the renormalized
Hamiltonian of the system contains an arbitrary parameter. 
Hamiltonians with different values of this parameter cannot be related via a
unitary transformation.
The limiting case of this system therefore admits a
one-parameter family of inequivalent quantizations. Finally, this model  
exhibits a type of T-duality symmetry. This feature can be used to
remove the quantization ambiguity only at the self-dual point.

In the case of a non-linear $\sigma$-model with a Wess-Zumino-Witten term a
similar result is obtained. The parameters of this model are however
constrained by topological considerations. However, in the limit when 
$S \rightarrow S_n$, the oscillatory modes of this system also have the same
behaviour as in the scalar field case. The renormalized Hamiltonian is
described only in terms of a finite number of degrees of freedom given by
the zero modes. It is also seen to contain an arbitrary parameter that 
can be related to one of the constants of the theory  when $v = 0$.
This observation however is not enough to fix a unique
 value of this parameter. We can hence say that this
system also admits a one-parameter family of inequivalent quantizations.

There seems to be a degree of universality associated with the results
obtained above. The suppression of the oscillatory modes in the limit of 
$v \rightarrow 1$ can be traced to Eqn. (2.11) and (3.51) for the scalar
field and the non-linear $\sigma$-model cases respectively. Such equations
would always occur whenever there is periodicity condition on the basic
variables of the theory concerned. We therefore conclude that the
suppression of the oscillatory modes would be a generic phenomenon in this
type of a scenario. This would in turn mean that as $S \rightarrow S_n$,
the resulting theory on the null surface would generically be described by a
finite number of degrees of freedom related to the zero modes of the system.

Both the models considered in this paper admits a one-parameter family of
inequivalent quantizations. We have however not found any general argument
supporing the universality of this phenomenon.

It would be interesting to perform similar analysis to more realistic models
of physical interest, e.g. boundary field theories on a black hole horizon
which is a null surface in space-time.
The analysis presented in this paper could be adapted to study the dynamics
of field theories on such a surface which is currently under investigation.

\newpage
\noindent {\bf\large Acknowledgments} 
\vskip 5mm
\noindent
We would like to thank A.P.Balachandran for suggesting the problem and
for his critical comments while the work was in progress.  
The second author would like also to thank 
Djamel Dou , Denjoe O'Connor and Garnik
 Alexanian for helpful discussions . The work of
B.Y was supported in part by the DOE under contract number
DE-FG02-85ER40231.

\vskip 5mm
\axn{Appendix}
\vskip 5mm
Here we give some of the identities necessary to derive Eqns. (3.34)
and (3.39).

To find the commutation relations among the conjugate momenta 
$ {\Pi}_a $ we proceed as follows. Consider the Jacobi identity
\begin{equation}
[[\Pi_a (\theta), \Pi_b ({\theta}^{'})], g ({\theta}^{''})]
= - [[\Pi_b ({\theta}^{'}), g ({\theta}^{''})], \Pi_a (\theta)]
  - [[g ({\theta}^{''}), \Pi_a (\theta)], \Pi_b ({\theta}^{'})].
\end{equation}
Using Eqns. (3.2) and (3.22), the above Jacobi identity gives \cite{bal}
\begin{equation}
[{\Pi}_a({\theta}),{\Pi}_b({\theta}^{'})]
=i f_{abc}{\Pi}_c({\theta}) {\delta}(\theta -{\theta}^{'})+F
\end{equation}
where $[F,g]=0$ . The fact that  $[F,g]=0$ implies that $F$ does 
not depend on $P_i$ but only on $g$ . 
Setting  $P_i=0$ in Eqn. (A.2) gives $F=0$. This proves Eqn. (3.34).

Next we sketch the steps leading to Eqn. (3.39).
First we prove the identity
\begin{equation}
f_{abd}N_d^{c}=-N_a^{d}{\frac{{\partial}N_b^{{c}}}{{\partial}{\xi}^{d}}}+
N_b^{d}{\frac{{\partial}N_a^{{c}}}{{\partial}{\xi}^{d}}}
\end{equation}
which will be used in the proof of the commutator . 
Using the definition (3.31) of ${\Pi}_a(\theta)$ we get
\begin{eqnarray}
[{\Pi}_a(\theta),{\Pi}_b({\theta}^{'})]
&=&[N_a^{c}(\theta)P_c(\theta), 
N_b^{d}({\theta}^{'})P_d({\theta}^{'})]\nonumber\\
&=&N_a^{c}(\theta)[P_c(\theta), N_b^{d}
({\theta}^{'})]P_d({\theta}^{'}) + 
N_b^{d}({\theta}^{'})[N_a^{c}({\theta}),
P_d({\theta}^{'})]P_c(\theta).\nonumber\\
\end{eqnarray}

From Eqn. (A.2) we get,
\begin{eqnarray}
[{\Pi}_a(\theta),{\Pi}_b({\theta}^{'})]
&=&if_{abc}{\delta}(\theta-{\theta}^{'}){\Pi}_c(\theta)\nonumber\\
\end{eqnarray}
Using Eqns. (A.4), (A.5) and (3.27), Eqn. (A.3) follows easily.

We are now ready to compute $[J^{0}_a(\theta),J^{0}_b({\theta}^{'}]$ . From
Eqn. (3.31) we get

\begin{eqnarray}
[\frac{-2iAR}{\gamma}J^{0}_a({\theta}),
\frac{-2iAR}{\gamma}J^{0}_b({\theta}^{'})]
&=&[{\Pi}_{a}(\theta),{\Pi}_{b}({\theta}^{'})]\nonumber\\
&+& [{\Pi}_{a}(\theta) , N_{b}^{i}({\theta}^{'}){\omega}_{ij}({\theta}^{'})
{\partial}_{{\theta}^{'}}{\xi}^{j}] + 
[N_{a}^{i}(\theta){\omega}_{ij}(\theta){\partial}_{\theta}{\xi}^j , 
{\Pi}_b({\theta}^{'}]\nonumber\\ 
&+&[N_{a}^{i}(\theta){\omega}_{ij}(\theta){\partial}_{\theta}{\xi}^{j}, 
N_{b}^{i}({\theta}^{'}){\omega}_{ij}({\theta}^{'})
{\partial}_{{\theta}^{'}}{\xi}^{j}].
\end{eqnarray}
The last commutator is  zero as it has 
 no time derivative . The first commutator is given by $(A.5)$ . The third
 commutator can be obtained from the second by
 interchanging $a$ with $b$ and ${\theta}$ with ${\theta}^{'}$ then
 putting an overall minus sign . Let us then compute the second comutator 
\begin{eqnarray}
[{\Pi}_{a}(\theta) , 
N_{b}^{i}({\theta}^{'}){\omega}_{ij}
({\theta}^{'}){\partial}_{{\theta}^{'}}{\xi}^{j}]&=&[{\Pi}_{a}(\theta) , 
N_{b}^{i}({\theta}^{'})]{\omega}_{ij}({\theta}^{'})
{\partial}_{{\theta}^{'}}{\xi}^{j} + 
N_{b}^{i}({\theta}^{'}){\omega}_{ij}({\theta}^{'})
[{\Pi}_{a}(\theta),{\partial}_{{\theta}^{'}}{\xi}^{j}]\nonumber\\
&+&N_{b}^{i}({\theta}^{'})[{\Pi}_{a}(\theta),
{\omega}_{ij}({\theta}^{'})]{\partial}_{{\theta}^{'}}{\xi}^{j}]\nonumber\\
&=&-i{\delta}(\theta -{\theta}^{'})
N_{a}^{c}\frac{{\partial}N_b^i}
{\partial{\xi}^c}{\omega}_{ij}
{\partial}_{\theta}{\xi}^{j} 
-i{\partial}_{{\theta}^{'}}{\delta}
(\theta -{\theta}^{'})N_{b}^{i}({\theta}^{'}) 
N_a^j(\theta) {\omega}_{ij}({\theta}^{'})\nonumber\\
&-&i{\delta}(\theta -{\theta}^{'})
N_{b}^{i}N_{a}^{c}\frac{{\partial}{\omega}_{ij}}{{\partial}{\xi}^c}
{\partial}_{\theta}{\xi}^j
\end{eqnarray}
where we have made use of Eqns. $(3.27)$ and $(3.31)$ . The sum
 of the second and the third commutator in Eqn. $(A.6)$ is then given by
\begin{eqnarray}
2+3&=&a + b +c\nonumber\\
\rm{where}~
~a&=&-i{\delta}(\theta -
{\theta}^{'})[N_{a}^{c}\frac{{\partial}N_b^i}
{\partial{\xi}^c}-N_{b}^{c}\frac{{\partial}N_a^i}{{\partial}{\xi}^c}]
{\omega}_{ij}{\partial}_{\theta}{\xi}^{j}\nonumber\\
b&=&-i{\partial}_{{\theta}^{'}}{\delta}
(\theta -{\theta}^{'})N_{b}^{i}({\theta}^{'}) 
N_a^j(\theta) {\omega}_{ij}({\theta}^{'})+
i{\partial}_{{\theta}}{\delta}(\theta -
{\theta}^{'})N_{a}^{i}({\theta}) N_b^j({\theta}^{'}) 
{\omega}_{ij}({\theta})\nonumber\\
c&=&-i{\delta}(\theta -{\theta}^{'})
[N_{b}^{i}N_a^c-N_a^iN_b^c]\frac{{\partial}{\omega}_{ij}}{{\partial{\xi}^c}}
\partial_{\theta}{\xi}^j.
\end{eqnarray}
By using $(A.3)$ we find that 
\begin{equation}
a=i{\delta}(\theta -
{\theta}^{'})f_{abc}N_c^i{\omega}_{ij}{\partial}_{\theta}{\xi}^j
\end{equation}
Careful manipulations with $b$ will give 
\begin{eqnarray}
b&=&-i{\partial}_{{\theta}^{'}}{\delta}
(\theta -{\theta}^{'})N_{b}^{i}({\theta}^{'}) 
N_a^j(\theta) {\omega}_{ij}({\theta}^{'})+
i{\partial}_{{\theta}}{\delta}(\theta -
{\theta}^{'})N_{a}^{i}({\theta}) N_b^j({\theta}^{'}) 
{\omega}_{ij}({\theta})\nonumber\\
&=&-iN_a^j(\theta)N_b^i({\theta}^{'})[{\partial}_{{\theta}^{'}}{\delta}
(\theta -{\theta}^{'}){\omega}_{ij}({\theta}^{'}) +
{\partial}_{{\theta}}{\delta}(\theta -
{\theta}^{'}){\omega}_{ij}({\theta})]\nonumber\\
&=&i{\delta}(\theta -
{\theta}^{'})N_a^j(\theta)N_b^i({\theta}^{'}){\partial}_{\theta}
{\omega}_{ij}\nonumber\\
&=&i{\delta}(\theta -
{\theta}^{'})N_a^cN_b^i 
\frac{{\partial}{\omega}_{ic}}{{\partial{\xi}^j}}\partial_{\theta}{\xi}^j.
\end{eqnarray}
Finally $c$ can be rewitten as
\begin{equation}
c=i{\delta}(\theta -
{\theta}^{'})N_a^c N_b^i
[\frac{{\partial}{\omega}_{cj}}{{\partial{\xi}^i}} + 
\frac{{\partial}{\omega}_{ji}}{{\partial{\xi}^c}}]\partial_{\theta}{\xi}^j.
\end{equation} 
Putting Eqns. $(A.9)$ , $(A.10)$ and $(A.11)$ together
 and using Eqns. $(3.19)$ and $(3.20)$, we get
\begin{eqnarray}
2+3&=&i{\delta}(\theta -
{\theta}^{'})f_{abc}N_c^i{\omega}_{ij}{\partial}_{\theta}{\xi}^j +
 i{\delta}(\theta -
{\theta}^{'})N_a^c N_b^i
[\frac{{\partial}{\omega}_{ic}}{{\partial{\xi}^j}} +
 \frac{{\partial}{\omega}_{cj}}{{\partial{\xi}^i}} + 
\frac{{\partial}{\omega}_{ji}}{{\partial{\xi}^c}}]
\partial_{\theta}{\xi}^j\nonumber\\
&=&i{\delta}(\theta -{\theta}^{'})
f_{abc}N_c^i{\omega}_{ij}{\partial}_{\theta}{\xi}^j + 
3iB {\delta}(\theta -{\theta}^{'})
N_a^cN_b^iTr[g^{-1}\frac{{\partial}{g}}
{\partial{\xi}^i},g^{-1}\frac{{\partial}{g}}
{\partial{\xi}^c}]g^{-1}g^{'}\nonumber\\
&=&i{\delta}(\theta -{\theta}^{'})
f_{abc}N_c^i{\omega}_{ij}{\partial}_{\theta}{\xi}^j + 
{\delta}(\theta -{\theta}^{'})3r^2Bf_{abc}J_a^{1}.
\end{eqnarray}
Using Eqns. $(A.12)$ 
and $(A.5)$ in Eqn. $(A.6)$ will lead to the commutation relations
\begin{equation}
[J^{0}_a({\theta}),J^{0}_b({\theta}^{'})]=-\frac{\gamma}{2AR} f_{abc} 
J^{0}_{c}(\theta) {\delta}(\theta - {\theta}^{'}) 
- \frac{3B{\gamma}^2}{4A^2} {\delta}({\theta} 
-{\theta}^{'}) f_{abc}J^{1}_{c}(\theta).
\end{equation}

\newpage
\bibliographystyle{unsrt}

\end{document}